\documentclass[aps,prl,showpacs,twocolumn]{revtex4}
\usepackage{epsfig}
\usepackage{amsmath}
\usepackage{graphicx}
\usepackage{amsfonts}
\usepackage{amssymb}

\input{tcilatex}

\begin{document}

\title{Optical realization of optimal unambiguous discrimination for pure
and mixed quantum states}
\author{Masoud Mohseni$^{1}$}
\author{Aephraim M. Steinberg$^{1}$}
\author{J\'{a}nos A. Bergou$^{2}$}
\affiliation{$^{1}$Department of Physics, University of Toronto, 60 St. George St.,
Toronto, Ontario, Canada, M5S 1A7}
\affiliation{$^{2}$Department of Physics and Astronomy, Hunter College of the City
University of New York, 695 Park Avenue, New York, NY 10021, USA}

\begin{abstract}
Quantum mechanics forbids deterministic discrimination among non-orthogonal
states. Nonetheless, the capability to distinguish nonorthogonal states
unambiguously is an important primitive in quantum information processing.
In this work, we experimentally implement generalized measurements in an
optical system and demonstrate the first optimal unambiguous discrimination
between three nonorthogonal states, with a success rate of 55\%, to be
compared with the 25\% maximum achievable using projective measurements.
Furthermore we present the first realization of unambiguous discrimination
between a pure state and a nonorthogonal mixed state.
\end{abstract}

\date{\today}
\maketitle

Quantum measurement theory poses fundamental limitations on the amount of
information that can be obtained about the state of a single quantum system.
Specifically, it is impossible to perfectly discriminate between two or more
nonorthogonal quantum states. However, by appropriately choosing a set of
measurements, \textit{nondeterministic} state discrimination is possible if
the system has been prepared in a member of a previously specified set of
nonorthogonal states. Quantum state discrimination plays an important role
in quantum information and quantum communications \cite{CheflesQSD} and is
at the heart of quantum cryptography protocols \cite{BB84}.

Several different strategies have been developed to accomplish
state-discrimination tasks. \textquotedblleft Minimum-error
discrimination\textquotedblright\ (MD) seeks a \textquotedblleft best
guess\textquotedblright\ on every trial, minimizing the rate of incorrect
guesses. Helstrom showed \cite{Helstrom} that, for MD of two states, the
optimal strategy can always be achieved by a (von Neumann) projective
measurement. \textquotedblleft Unambiguous state
discrimination,\textquotedblright\ (UD), on the other hand, seeks to
determine with certainty which state the system was in. This can only be
done on some fraction of the trials, the others being termed
\textquotedblleft inconclusive,\textquotedblright\ and the optimal UD
success rate \textit{cannot} always be achieved with projective
measurements. For the case of two pure nonorthogonal quantum states with
equal \textit{a priori} probability, the maximum probability of success was
derived by Ivanovic, Dieks and Peres \cite{IvanDieksPeres}. Clearly, these
two strategies may be regarded as limiting cases of a more general approach
with both a finite inconclusive rate and a finite error rate \cite{tradeoff}%
. Of course, in any experimentally realistic situation, even an ideal
\textquotedblleft unambiguous\textquotedblright\ discrimination strategy
will not be error-free. It is of great importance to understand the
theoretical and practical limitations, since state discrimination is part of
quantum key distribution protocols, and because the effect of a hypothetical
eavesdropper's attack on such cryptosystems requires knowledge of the
maximum information she could extract.

As for the experimental state-of-the-art, for two nonorthogonal states UD
was qualitatively demonstrated by Huttner \textit{et al.} \cite{Huttner},
while MD was demonstrated by Barnett and Riis \cite{firstex}. Clarke \textit{%
et al.} demonstrated UD quantitatively for a pair of nonorthogonal states
\cite{Clarke1}. In a second experiment, using highly symmetric trine and
tetrad states (linearly dependent states in two dimensions), they performed
optimum MD and also a projective measurement which could indicate that a
qubit was not prepared in one out of three or four possibilities \cite%
{Clarke2}. Thus, all of the previous experiments have been limited to
nonorthogonal pure states of a qubit in a two-dimensional Hilbert space. In
the present work, by constructing a multirail optical interferometer
enabling us to perform a large class of generalized measurements, we extend
these results to higher-dimensional Hilbert spaces with no restriction on
the symmetry of the states and explicitly demonstrate that we can achieve a
significantly higher success rate than any projective measurement. We
perform the first optimal unambiguous state discrimination for three
nonorthogonal states, showing that the experimental success rate may be as
much as twice as high as for any von Neumann measurement scheme. We also
demonstrate the first optimal quantum state \textquotedblleft
filtering,\textquotedblright\ discriminating between two \textit{subsets} of
a set of three nonorthogonal states. This is equivalent to discrimination
between a pure and a mixed state \cite{HB,Terry}.

Unambiguous state discrimination between $N$ states has $N+1$ outcomes: the $%
N$ possible conclusive results, and the inconclusive result. Since no
projective measurement in an $N$-dimensional Hilbert space can have more
than $N$ outcomes, generalized measurements are required. Generalized
measurements (or positive-operator valued measures, POVMs) provide the most
general means of transforming the state of a quantum system \cite%
{Helstrom,Nielsen}. POVMs can be implemented by embedding the system into a
larger Hilbert space and unitarily entangling it with the extra degrees of
freedom (ancilla) \cite{syssubsys}. Postselection (projective measurement)
of the ancilla induces an effective non-unitary transformation of the
original system. By an appropriate design of the entangling unitary, this
effective non-unitary transformation can turn an initially nonorthogonal set
of states into a set of orthogonal states with a finite probability of
success. The optimum strategy is the one that maximizes the average
probability of success for this procedure. The generalization of \cite%
{IvanDieksPeres} to more than two states was developed by several groups
\cite{linearindep,CheflesBarNSD}.

The problem of distinguishing among two \textit{subsets} of a set of $N$
nonorthogonal quantum states has been termed \textquotedblleft quantum state
filtering" for the case when one subset contains $1$ state and the other $N-1
$ \cite{HB}. Unambiguous filtering has been studied by Bergou \textit{et al.}
for $N=3$ in \cite{BergouSF1} and for arbitrary $N$ in \cite{BHH}. Quantum
state filtering can also be interpreted as unambiguous discrimination
between two mixed states. Let us consider the case of three nonorthogonal
states, $\{\left\vert \psi _{1}\right\rangle, \left\vert \psi
_{2}\right\rangle, \left\vert \psi _{3}\right\rangle \}$ with \textit{a
priori} probabilities of $\eta _{1},\eta _{2}$ and $\eta _{3}$. Filtering is
the optimal strategy that can distinguish the state $\left\vert \psi
_{1}\right\rangle $ from the subset $\{\left\vert \psi_{2}\right\rangle ,
\left\vert \psi _{3}\right\rangle \}$. This is equivalent to discriminating
the pure state $\left\vert \psi _{1}\right\rangle \left\langle \psi
_{1}\right\vert $, with \textit{a priori} probability of $\eta _{1}$, from
the mixed state $(\eta _{2}\left\vert \psi _{2}\right\rangle \left\langle
\psi _{2}\right\vert +\eta _{3}\left\vert \psi_{3}\right\rangle \left\langle
\psi _{3}\right\vert)/(\eta _{2}+\eta _{3}) $, with \textit{a priori}
probability of $\eta_{2}+\eta _{3}$.

Here, we present our experimental data for implementing the optimal POVM
both for the case of full UD of and filtering from $N=3$ linearly
independent nonorthogonal states. In our experiment the generalized
measurement was realized by utilizing linear optical elements and
photodetectors, based on the proposal of \cite{BergouSD}. This can be
accomplished by a single-photon representation of the initial states and
output states, a multirail optical network for performing the unitary
transformation \cite{Reck,ZukowskiZeilinger}, and photodetectors at each
output port to carry out the required nonunitary transformation. The $N+1$%
-dimensional unitary operation can be implemented by an appropriate
multi-path optical interferometer \cite{Reck,BergouSD}. For the case of
three nonorthogonal states, $\{\left\vert \psi _{1}\right\rangle,\left\vert
\psi _{2}\right\rangle ,\left\vert \psi _{3}\right\rangle\}$, living in a
3-D Hilbert space, an eight-port optical interferometer was constructed to
perform transformations in the 4-D system + ancilla space. All beam
splitters in this interferometer were designed (using a combination of
polarizing beam splitters and waveplates) to have variable reflectivity, so
that the appropriate interferometer could be implemented for any desired
discrimination problem (see Fig. 1).%

\begin{figure}[ht]
\epsfig{file=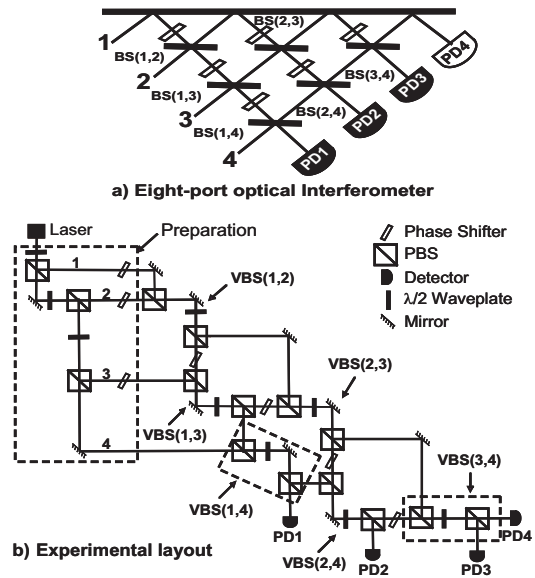,height=3in}
\caption{(a) Eight-port optical interferometer: Suitable beam splitters are
placed at each crossing of two optical rails to realize any desired unitary
transformation on the input states. A detection in rail 4 corresponds to an
inconclusive result and a detection in rails 1 to 3 corresponds to states $|%
\protect\psi _{1}\rangle$ to $|\protect\psi _{3}\rangle$. (b) Experimental
layout: This interferometer can perform various desired generalized
measurement by doing arbitrary unitary operations in four-dimensional
Hilbert space and projective measurements at output ports 1 to 4. The
variable beam splitters (VBS) realize the corresponding beam splitters in
(a) for arbitrary reflectivity and transmissions. Photodiodes PD1 to PD4
detect the photons at the output ports 1 to 4.}
\label{Fig1}
\end{figure}

By using beam splitters to send one photon into some linear superposition of
the first three rails, we can generate arbitrary quantum states in this
three-dimensional Hilbert space, represented as $\left\vert \psi
\right\rangle _{in}=\sum_{j=1}^{3}c_{j}\hat{a}_{j}^{\dagger }\left\vert
0\right\rangle $, where $\sum_{j=1}^{3}\left\vert c_{j}\right\vert ^{2}=1$,
and $\hat{a}_{j}^{\dagger }$ is the creation operator for the $j^{th}$
optical rail. Note that the fourth rail, which acts as the ancilla, never
contains a photon. The interferometer is designed to perform the unitary
operation $U$ which optimizes state discrimination. It maps the input field
operators $\hat{a}_{k}^{\dagger }$ into output field operators as $\hat{a}%
_{k}^{\dagger }=\sum_{j=1}^{4}U_{jk}\hat{a}_{j_{out}}^{\dagger }$, such that
the initial state evolves into $\mbox{$|\psi\protect\rangle_{out}$}%
=\sum_{j=1}^{j=4}\sum_{k=1}^{k=3}U_{jk}c_{k}\hat{a}_{j_{out}}^{\dagger }%
\mbox{$|0\protect\rangle$}$. A photon in mode 4 now indicates an
inconclusive result. On the other hand, a photon in mode $1,2$ or
$3$ unambiguously indicates that the initial state was $\left\vert
\psi _{1}\right\rangle ,\left\vert \psi _{2}\right\rangle $ or
$\left\vert \psi _{3}\right\rangle $, respectively.

The actual experimental setup is shown in Fig. 1(b). To demonstrate
unambiguous discrimination, and characterize the success and error rates of
our setup, we performed the experiment using a large ensemble of identically
prepared photons from a diode laser operating at 780 nm. The nonorthogonal
states were prepared by using an arrangement of polarizing beam splitters
(PBS), half-wave plates and phase shifters. A 1-mm-thick glass slide, at an
angle of 55$^{\circ }$ to the incident beam, was used as a phase shifter. At
this angle a differential rotation of 0.05$^{\circ }$ of the phase shifter
produces a $\pi $ phase shift and causes a beam displacement of less than a
micron. Each nonorthogonal state consisted of a different superposition of
light in rails 1-3, with the relative field amplitude being adjusted to
generate the appropriate coefficients $c_{j}$. Rail 4 contained the vacuum
for all input states \cite{footnote}. We designed a variable beam splitter
(VBS) that could be placed at each crossing of the beams in Fig. 1(a), in
order to perform arbitrary 4-D unitaries. The VBS consists of three
half-wave plates and two polarizing beam splitters (PBS) \cite{masoud}. The
polarizing beam splitters are used to convert information between spatial
and polarization degrees of freedom, such that instead of arbitrary coupling
between two spatial modes, we only need to implement arbitrary coupling
between two polarizations, easily accomplished using waveplates. The setup
was designed such that in all interferometers, all the spatial path lengths
are always balanced.

To perform the discrimination or filtering for a specific set of three
nonorthogonal states, with equal \textit{a priori} probabilities, the
optimal success probability and output states were calculated following the
method in \cite{BergouSF1}. Using the input and output states in the larger
Hilbert space, the corresponding unitary transformation was then calculated
and factorized into a sequence of beam splitter transformations. By rotation
of the half-wave plates in each of the VBS's the desired transmission and
reflectivity were achieved. The experiments were carried out by preparing
one of the three nonorthogonal states at a time and measuring the current at
the photodiodes PD1 through PD4. For obtaining the probability of an
individual photon reaching each detector, the signals at these detectors
were normalized to their sum.

In order to demonstrate state discrimination with this experimental setup we
examined the set of three nonorthogonal pure states $|\psi _{1}\rangle =(%
\sqrt{2/3},0,1/\sqrt{3})$ and $|\psi _{2,3}\rangle =(0,\pm 1/\sqrt{3},\sqrt{%
2/3})$. The optimal output states, in the total Hilbert space, are found to
be: $\left\vert \psi _{1}\right\rangle _{out}=%
\begin{pmatrix}
1/\sqrt{3}, & 0, & 0, & \sqrt{2/3}%
\end{pmatrix}%
$, $\left\vert \psi _{2}\right\rangle _{out}=%
\begin{pmatrix}
0, & \sqrt{2/3}, & 0, & \sqrt{1/3}%
\end{pmatrix}%
$ and $\left\vert \psi _{3}\right\rangle _{out}=%
\begin{pmatrix}
0, & 0, & \sqrt{2/3}, & \sqrt{1/3}%
\end{pmatrix}%
$. The desired unitary transformation was achieved by using two VBSs with
the transmission coefficients $t_{14}=t_{34}=1/\sqrt{2}$ (and $t=1$ for the
rest of the VBSs), and an additional 50/50 beam splitter to couple output
rails 2 and 3.

The experimental results are shown in Fig. 2.
\begin{figure}[ht]
\epsfig{file=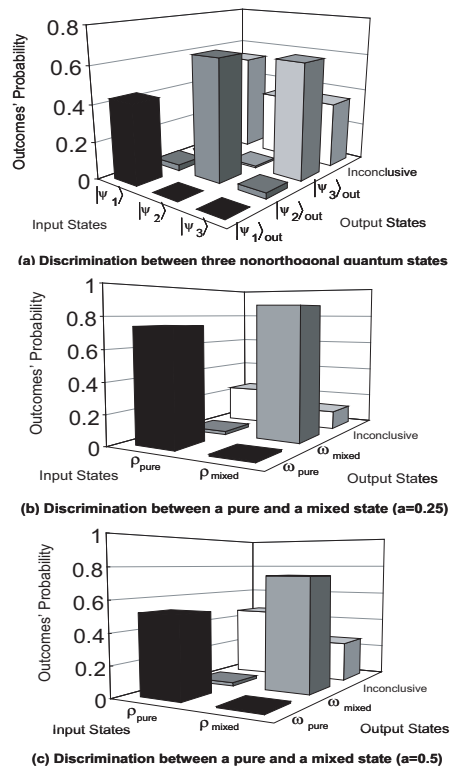,height=4in}
\caption{Experimental data: The results of state discrimination, and of
state filtering for the cases of a=0.25 and a=0.5 respectively, are
presented in parts (a), (b), and (c). Each row corresponds to preparation of
a pure or a mixed quantum state. The last column in each figure represents
the inconclusive outcomes. The diagonal and off-diagonal elements of other
columns represent the successful and erroneous detections respectively. The
probability of each outcome is a measure of the fraction of photons reaching
the corresponding detector.}
\label{Fig2}
\end{figure}
Figure 2(a) pertains to the case when all states are discriminated. The
average probability of success was measured to be 54.5\%. The probability of
obtaining an erroneous result was about 3\%. These errors were mainly the
result of imperfect visibility (due to imperfect alignment and angular
uncertainty in the waveplate settings), drift and uncertainty in phase
adjustment, where a $12^{\circ }$ phase error on one beam corresponds to a
2\% error rate. For this set the optimal POVM is predicted to yield
conclusive outcomes 55.6\% of the time. By comparison, \textit{any}
projection valued measurement (PVM) strategy has a success probability of
less than 33.3\%. This is because the only way a PVM can \textit{guarantee}
that we had $|\psi_{i}\rangle$ is to project onto the unique state
orthogonal to all input vectors $|\psi_{j\neq i}\rangle$. In this case, all
three such vectors are nonorthogonal, so no orthonormal set of projectors
can include more than one of them; no more than one of the three states can
be unambiguously distinguished. Given equal \textit{a priori} probabilities,
this means that PVMs can succeed no more than 33.3\% of the time. In fact,
for our input states the optimum PVM is the one that picks out $%
|\psi_{2}\rangle$ (or $|\psi_{3}\rangle$, their success probabilities being
equal), with a 25.4\% probability of success. In the above example we have
shown an improvement of more than a factor of 2 over any possible projective
measurement.

For quantum state filtering we considered a family of three-state sets, $%
\left\vert \psi_{1}\right\rangle = (\sqrt{1-a^{2}}, 0, a)$ and $\left\vert
\psi_{2,3}\right\rangle = (0, \pm 1/\sqrt{2}, 1/\sqrt{2})$, characterized by
the real parameter $a>0$. The goal was to unambiguously distinguish $%
\left\vert \psi_{1}\right\rangle$ from the other two states $\left\{
\left\vert \psi _{2}\right\rangle ,\left\vert \psi _{3}\right\rangle \right\}
$, each of which has an overlap of $\frac{a}{\sqrt{2}}$ with $\left\vert
\psi _{1}\right\rangle$. The optimal output states are $|\psi
_{1}\rangle_{out} = (\sqrt{1-a}, 0, 0, \sqrt{a})$ and $|\psi_{2,3}%
\rangle_{out} = (0, \pm1/\sqrt{2}, \sqrt{(1-a)/2}, \sqrt{a/2})$. The unitary
transformation for optimal filtering was achieved by beam splitters with
parameters $\ t_{14}=1/\sqrt{1+a}$ and $t_{34}=\sqrt{1-a}$, and $t=1$ for
the other VBS's in the setup. For these sets of states, the optimal success
probabilities PVMs and POVMs are $\left(2-a^{2}\right) /3$ and $1-(2a)/3$,
respectively.

The experiments were performed for $a=0.25$ and $a=0.5$. Fig. 2(b) pertains
to the case $a=0.25$. The average probability of success for discriminating
the state $|\psi_{1}\rangle$ from $\left\{\left\vert \psi
_{2,3}\right\rangle\right\}$ was measured to be about 82\%, consistent with
the theoretical prediction of 83.3\%. There was an error rate of about
1.7\%. The advantage of the POVM measurements over error-free projective
measurements was found to be about 17.4\% for this case. For $a=0.5$, Fig.
2(c), the probability of success was found to be about 66\%, with an error
probability of less than 1.3\%. This can be compared with the theoretical
predictions of 66.6\%. In this case the advantage over PVMs reduces to
7.7\%. As we argued above, these filtering experiments are equivalent to
discrimination between the pure state $\rho_{1}=\left\vert\psi
_{1}\right\rangle \left\langle \psi _{1}\right\vert $, with $\eta_{1}=1/3$,
and the mixed state $\rho_{23}=(\left\vert \psi _{2}\right\rangle
\left\langle \psi_{2}\right\vert +\left\vert
\psi_{3}\right\rangle\left\langle \psi _{3}\right\vert)/2 $, with $%
\eta_{23}=2/3$. The experimental results are summarized in Table 1.
\begin{table}[h!]
\caption{Experimental and theoretical success probabilities of POVMs vs. the
optimal PVM for State Filtering (SF, for a= 0.25 and a=0.5) and State
Discrimination (SD).}%
\begin{tabular}{||c||c|c|c||}
\hline
& $SF(a=0.25)$ & $SF(a=0.50)$ & $SD$ \\ \hline\hline
$POVM_{exp}$ & 82\% & 66\% & 54.5\% \\ \hline
$POVM_{th}$ & 83.3\% & 66.6\% & 55.6\% \\ \hline
$PVM_{th}$ & 64.6\% & 58.3\% & 25.4\% \\ \hline
\end{tabular}
\newline
\end{table}

In conclusion, we have presented the first explicit experimental
demonstration that POVMs can achieve a lower inconclusive rate than any
projective measurements for unambiguous discrimination between nonorthogonal
states, using an optical interferometer to implement arbitrary unitary
operations in a 4-dimensional Hilbert space. We have demonstrated the first
optimal unambiguous discrimination between three linearly-independent
nonorthogonal pure states, as well as the first experimental realization of
unambiguous discrimination between a pure and a mixed quantum state
(\textquotedblleft quantum state filtering\textquotedblright ). A
significant advantage of generalized measurement over projective measurement
was observed. Unambiguous state discrimination plays an important role in
the field of quantum information processing and has applications to quantum
cryptography \cite{BB84}, quantum cloning \cite{cloning}, quantum state
separation \cite{stateseparation} and entanglement concentration \cite%
{linearindep,CheflesQSD}. Some quantum information tasks are likely to take
advantage of 3- or higher-dimensional Hilbert spaces \cite{Zeilinger}, where
the advantage of POVMs becomes increasingly significant, as observed here.
We believe that generalized optical networks like the one demonstrated here
will be of use for a wide variety of small-scale quantum information tasks
\cite{CerfKwiat,masoud}, and prove particularly important for devices such
as repeaters and cloners in quantum communications systems.

This work was supported by the DARPA-QuIST program (AFOSR agreement No.
F49620-01-1-0468), NSERC, and Photonics Research Ontario. The research of JB
was supported by ONR. We would like to thank Jeffrey Lundeen, Kevin Resch,
Mark Hillery, Ulrike Herzog and Yuqing Sun for many helpful discussions.


\begin{thebibliography}{99}
\bibitem{CheflesQSD} A. Chefles, Contemp. Phys. \textbf{41}, 401  (2000).

\bibitem{BB84} C. H. Bennett and G. Brassard, in \textit{Proc. of the IEEE
International Conference on Computers, Systems and Signal Processing} (IEEE,
New York, 1984), p. 175.

\bibitem{Helstrom} C. W. Helstrom, \textit{Quantum Detection and Estimation
Theory} (Academic Press, New York, 1976).

\bibitem{IvanDieksPeres} I. D. Ivanovic, Phys. Lett. A \textbf{123}, 257
(1987); D. Dieks,\ Phys. Lett. A \textbf{126}, 303  (1988); A. Peres, Phys.
Lett. A \textbf{128}, 19 (1988).

\bibitem{tradeoff} A. Chefles and S. M. Barnett,  J. Mod. Opt. \textbf{45},
1295 (1998).

\bibitem{Huttner} B. Huttner \textit{et al.}, \pra \textbf{54}, 3783  (1996).

\bibitem{firstex} S. M. Barnett and E. Riis, J. Mod. Opt. \textbf{44},  1061
(1997).

\bibitem{Clarke1} R. B. M. Clarke \textit{et al.}, \pra\textbf{63},
040305(R) (2001).

\bibitem{Clarke2} R. B. M. Clarke \textit{et al.}, \pra {\bf 64}, 012303
(2001).

\bibitem{HB} U. Herzog and J. A. Bergou, \pra {\bf 65}, 050305(R)  (2002).

\bibitem{Terry} T. Rudolph, R. W. Spekkens, and P. S. Turner, Phys. Rev. A
\textbf{68}, 010301(R) (2003).

\bibitem{Nielsen} M. A. Nielsen and I. L. Chuang, \textit{Quantum
Computation and Quantum Information} (Cambridge University Press  (2000).

\bibitem{syssubsys} M. A. Neumark, Izv. Akad. Nauk. SSSR, Ser. Mat. \textbf{%
4 53}, 277 (1940).

\bibitem{linearindep} A. Chefles, Phys. Lett. A \textbf{239}, 339  (1998).

\bibitem{CheflesBarNSD} A. Chefles and S. M. Barnett, Phys. Lett. A  \textbf{%
250}, 223 (1998); C.W. Zhang, C.F. Li, and G.C. Guo, Phys. Lett. A  \textbf{%
261}, 25 (1999).

\bibitem{BergouSF1} Y. Sun, J. A. Bergou, and M. Hillery, \pra
\textbf{66}, 032315 (2002).

\bibitem{BHH} J. A. Bergou, U. Herzog, and M. Hillery, \prl {\bf 90}, 257901
(2003).

\bibitem{BergouSD} Y. Sun, M. Hillery, and J. A. Bergou, \pra \textbf{64},
022311 (2001) for N=3; J. A. Bergou, M. Hillery, and Y. Sun, J. Mod. Opt.
\textbf{47}, 487 (2000) for N=2.

\bibitem{Reck} M. Reck, \textit{et al.}, \prl \textbf{73}, 58 (1994).

\bibitem{ZukowskiZeilinger} M. Zukowski,\textit{\ et al.}, \pra\textbf{55},
2564 (1997).

\bibitem{footnote} For example, to prepare the state $|\psi \rangle =(1/%
\sqrt{2},1/\sqrt{2},0)$, we align the first and second HWP's at 22.5$^{\circ
}$ and 0$^{\circ }$, so the first PBS acts as a 50/50 beam splitter and the
second has transmission 1.

\bibitem{masoud} M. Mohseni, J. S. Lundeen, K. J. Resch, and  A. M.
Steinberg, \prl \textbf{91}, 187903 (2003).

\bibitem{cloning} L.-M. Duan and G.-C. Guo, \prl \textbf{80}, 4999  (1998).

\bibitem{stateseparation} A. Chefles and S. M. Barnett, J. Phys. A  \textbf{%
31}, 10097 (1998).

\bibitem{Zeilinger} D. Kaszlikowski, \textit{et al.}, \prl\textbf{85},  4418
(2000); T. Durt, \textit{et al.}, \pra {\bf 67}, 012311  (2003).

\bibitem{CerfKwiat} N. J. Cerf, C. Adami, P.G. Kwiat, \pra \textbf{\ 57},
R1477 (1998).
\end{thebibliography}
\end{document}